\documentclass[journal]{IEEEtran}

\usepackage[dvips]{color}
\usepackage{graphicx}
\usepackage{amsmath}

\begin{document}

\title{Intelligent Interference Exploitation for Heterogeneous Cellular Networks against Eavesdropping}

\markboth{IEEE Journal on Selected Areas in Communications (ACCEPTED TO APPEAR)}%
{Yulong Zou: Intelligent Interference Exploitation for Heterogeneous Cellular Networks against Eavesdropping}

\author{Yulong~Zou,~\IEEEmembership{Senior Member,~IEEE}

\thanks{Manuscript received July 31, 2017; revised July 8, 2018. This work was partially supported by the National Natural Science Foundation of China (Grant Nos. 61522109, 61631020, 61671253 and 91738201), the Natural Science Foundation of Jiangsu Province (Grant Nos. BK20150040 and BK20171446), and the Key Project of Natural Science Research of Higher Education Institutions of Jiangsu Province (No. 15KJA510003).}

\thanks{Y. Zou is with the School of Telecommunications and Information Engineering, Nanjing University of Posts and Telecommunications, Nanjing, China. (Email: \{yulong.zou\}@njupt.edu.cn)}

}

\maketitle

\begin{abstract}
This paper explores the co-existence of a macro cell and a small cell for heterogeneous cellular networks, where a macro base station (MBS) and small base station (SBS) transmit to respective macro user (MU) and small user (SU) through their shared spectrum in the face of a common eavesdropper. We consider two spectrum sharing mechanisms, namely the overlay spectrum sharing (OSS) and underlay spectrum sharing (USS). In the OSS, MBS and SBS take turns to access their shared spectrum. By contrast, the USS allows MBS and SBS to simultaneously transmit over the shared spectrum with the aid of power control for limiting their mutual interference, thus called interference-limited USS (IL-USS). In order to take advantage of mutual interference in confusing the eavesdropper without causing adverse effect on the MU, we propose an interference-canceled USS (IC-USS) scheme, where a sophisticatedly-designed signal is emitted at MBS to cancel out the interference received at MU, which is also beneficial in terms of defending the common eavesdropper. Closed-form expressions of overall outage probability and intercept probability are derived for OSS, IL-USS and IC-USS schemes by taking into account both MBS-MU and SBS-SU transmissions. The secrecy diversity analysis is also carried out by characterizing an asymptotic behavior of the overall outage probability with a given intercept probability in the high signal-to-noise ratio region. It is shown that the secrecy diversity gains of conventional OSS and IL-USS are zero, whereas the proposed IC-USS achieves a higher secrecy diversity gain of one. This implies that with an arbitrarily low overall intercept probability, the conventional OSS and IL-USS methods converge to their respective outage probability floors, however the proposed IC-USS scheme can make the overall outage probability asymptotically decrease to zero by simply increasing the transmit power. Additionally, numerical results demonstrate an obvious advantage of the proposed IC-USS over OSS and IL-USS against eavesdropping.

\end{abstract}


\section{Introduction}

\IEEEPARstart{R}{ecently}, with an explosive growth of wireless traffic, an increasing research attention from academia and industry has been paid to the development of future ultra-high data-rate mobile communications systems e.g. 5G and beyond [1]-[3]. As a consequence, heterogeneous cellular networks consisting of a number of small cells (e.g. pico cells and femto cells) densely deployed in a macro cell, are emerging as a means of improving the spectrum efficiency and data rate of wireless systems [4], [5]. To be specific, in heterogeneous cellular networks, a macro base station (MBS) and a small base station (SBS) are allowed to share the same spectrum and transmit their confidential information to respective users over the shared spectrum, leading to a higher spectrum utilization [6], [7]. However, mutual interference between the macro cell and underlaying small cells may arise and severely degrades the quality-of-service (QoS) of heterogeneous cellular networks. To this end, power control and allocation [8]-[10] as well as interference alignment and management [11], [12] have been widely studied to suppress the mutual interference in heterogeneous cellular networks for system performance improvement.

\indent Meanwhile, due to the broadcast nature and inherent openness of wireless communications, an eavesdropper may wiretap both the macro-cell and small-cell transmissions, as long as it lies in the coverage of heterogeneous cellular networks. In order to defend against eavesdropping attacks, cryptographic methods are generally employed in cellular networks to guarantee the transmission confidentiality at the expense of extra computational complexity and latency resulted from the secret key management and encryption/decryption algorithms [13]. Alternatively, physical-layer security emerges as a promising paradigm to achieve the perfect secrecy by taking full advantage of physical characteristics of wireless channels [14], [15]. In [16], Wyner first proved that if the wiretap channel spanning from a source to an eavesdropper is a degraded version of the main channel spanning from the source to its desired receiver, a perfect secrecy can be achieved without any confidential information leakage to the eavesdropper. Later on, the authors of [17] introduced a notion of secrecy capacity shown as the difference between the capacity of main channel and that of wiretap channel, which is severely degraded in wireless fading environments. As a consequence, extensive efforts were devoted to examining the use of multiple antennas [18]-[21], multiuser scheduling [22] and cooperative relays [23]-[25] for enhancing the secrecy capacity for wireless communications.

\indent Also, there is a significant amount of research work focused on physical-layer security for heterogeneous spectrum-sharing networks [26]-[28], including cognitive radio (CR) networks and device-to-device (D2D) underlay cellular networks. More specifically, CR enables an unlicensed wireless network (often referred to as secondary network) to access licensed spectrum resources that are not occupied by a primary network [26], [27], where the secondary network has a lower priority than the primary network in accessing the licensed spectrum. By contrast, in D2D underlaying cellular networks [28], D2D communication means a direct link between two cellular users without traversing a base station (BS), which is allowed to share the same spectrum with its underlaying cellular links between BS and associated users. In [29] and [30], the secrecy capacity and outage performance of secondary transmissions were studied with a QoS guarantee of primary transmissions for CR networks. The security-reliability tradeoff (SRT) for CR systems was investigated in [31] and two relay selection schemes were proposed for SRT improvement, namely the single-relay selection and multi-relay selection. Additionally, physical-layer security for D2D underlaying cellular networks was studied in [32] by exploiting D2D scheduling to guarantee the secrecy performance of cellular communications. The authors of [33] further examined the maximization of secrecy capacity through power allocation between D2D and cellular links for D2D underlaying cellular networks.

In this paper, we explore physical-layer security for a heterogeneous spectrum-sharing cellular network consisting of a macro cell and a small cell, where a MBS and an SBS transmit to their respective macro user (MU) and small user (SU) and an eavesdropper intends to wiretap both MBS-MU and SBS-SU transmissions. The overlay spectrum sharing (OSS) and underlay spectrum sharing (USS) are considered for MBS-MU and SBS-MU links to access the same spectrum. The main contributions of this paper are summarized as follows. First, we propose an interference-canceled USS (IC-USS) scheme to take full advantage of mutual interference between the macro cell and small cell in confusing the common eavesdropper. To be specific, the USS enables both MBS and SBS to transmit their messages simultaneously, and thus leads to the mutual interference which has an adverse effect on decoding source messages at legitimate MU and SU, but is also beneficial to confuse the eavesdropper. To alleviate the mutual interference, power control is utilized in conventional interference-limited USS (IL-USS) scheme. By contrast, in our IC-USS scheme, a special signal is sophisticatedly designed for interference cancelation to reduce the adverse impact of mutual interference on legitimate MU and SU, which simultaneously generates certain interference to the undesired eavesdropper. Second, we derive closed-form expressions of overall outage probability and intercept probability for both MBS-MU and SBS-SU transmissions relying on the conventional OSS and IL-USS as well as our IC-USS schemes. Finally, the secrecy diversity analysis of OSS, IL-USS and IC-USS schemes is conducted by characterizing an asymptotic behavior of the overall outage probability with a given intercept probability in the high signal-to-noise ratio (SNR) region.

The reminder of this paper are organized as follows. In Section II, we provide the system model of heterogeneous spectrum-sharing cellular networks and propose OSS, IL-USS and IC-USS schemes. Next, we analyze the overall outage probability and intercept probability for OSS, IL-USS and IC-USS in Section III, followed by Section IV, where the secrecy diversity analysis is carried out. Then, we present numerical results of the overall outage probability and intercept probability in Section V. Finally, Section VI gives some concluding remarks.

\section{Spectrum Sharing for Heterogeneous Wireless Networks}
In this section, we first present the system model of a heterogeneous macro-cell and small-cell wireless system, where an eavesdropper is assumed to tap any active transmissions in both the macro cell and small cell. Then, we consider two different spectrum sharing mechanisms for the heterogeneous wireless system, namely the overlay spectrum sharing (OSS) and underlay spectrum sharing (USS).

\subsection{System Model}
As shown in Fig. 1, we consider a heterogeneous wireless system consisting of a macro cell and a small cell, where a macro base station (MBS) and small base station (SBS) transmit their confidential messages to respective intended users, called macro user (MU) and small user (SU), respectively. {{It needs to be pointed out that MBS and SBS are connected to a core network, e.g., a mobility management entity (MME) in the long term evolution (LTE) or a mobile switch center (MSC) in global system for mobile communication (GSM), through which reliable information exchange can be achieved between MBS and SBS.}} Presently, such a heterogeneous cellular architecture of embedding small base stations (e.g. pico/femto base stations) into a macro cell has been adopted in long term evolution-advanced (LTE-A) network [4], which is capable of substantially improving spectral efficiency and attractive to future evolved wireless networks.

In Fig. 1, a passive eavesdropper (E) is considered to tap MBS-MU and SBS-SU transmissions and assumed to know everything about the confidential transmissions (e.g., encryption/decryption algorithms and secret keys) except the source messages. Since the eavesdropper is passive, the channel state information (CSI) of the eavesdropper is assumed to be unavailable in this paper. Although only a macro cell, a small cell and an eavesdropper are included in the system model, a possible future extension may be considered for a large-scale heterogeneous network consisting of multiple MUs, SUs and eavesdroppers with the aid of user scheduling [22] and stochastic geometry [34], [35]. Moreover, in the heterogeneous wireless system of Fig. 1, the macro cell and small cell share the same spectrum resources. Throughout this paper, we consider two different spectrum sharing mechanisms i.e. OSS and USS.
\begin{figure}
\centering
\includegraphics[scale=0.45]{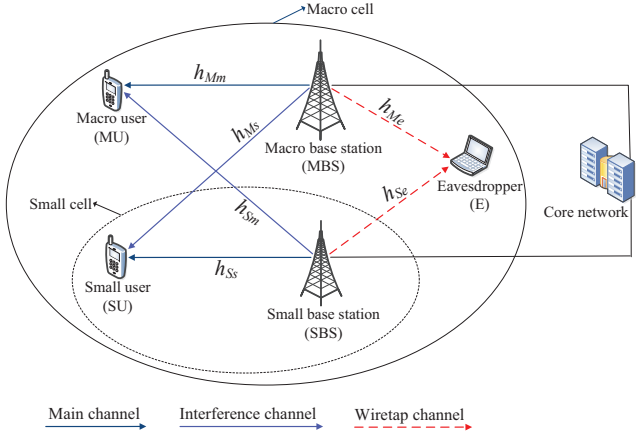}
\caption{A heterogeneous cellular network consisting of a macro cell and a small cell in the presence of a common eavesdropper.}
\label{fig 1}
\end{figure}

To be specific, in the OSS mechanism, a given spectrum band is divided into two orthogonal parts, which are allocated to the macro cell and small cell, respectively. In this manner, MBS and SBS transmit their messages over two orthogonal sub-bands without mutual interference. By contrast, the USS strategy allows MBS and SBS to transmit over the same spectrum band simultaneously with the aid of power control to limit the mutual interference level for the QoS guarantee. For national convenience, let $P_M$ and $P_S$ respectively denote transmit powers of MBS and SBS, where subscripts $M$ and $S$ represent MBS and SBS, respectively. Moreover, data rates of the MBS-MU and SBS-SU transmissions are denoted by $R^o_M$ and $R^o_S$, respectively. Throughout this paper, all the wireless links between any two nodes of Fig. 1 are modeled as Rayleigh fading. In addition, any receiver of Fig. 1 is assumed to encounter the zero-mean additive white Gaussian noise (AWGN) with a variance of ${N_0}$.

\subsection{Conventional OSS}
In this section, we consider the conventional OSS scheme as a baseline. As aforementioned, in the OSS scheme, a given spectrum band is first divided into two orthogonal sub-bands which are then assigned to the macro cell and small cell, respectively. As a result, no interference occurs between the macro cell and small cell, when MBS and SBS transmit to their respective users. We consider that a fraction of the total spectrum $\alpha$ is assigned to MBS and the remaining spectrum is allocated to SBS, wherein $0 \le \alpha \le 1$. We consider that MBS and SBS transmit their signals ${x_M}$ and ${x_S}$ at the power of ${P_M}$ and ${P_S}$, respectively, where $E(|x_M|^2)=1$, $E(|x_S|^2)=1$, and $E(\cdot)$ denotes an expectation operator. Thus, the received signal at MU can be expressed as
\begin{equation}\label{equa1}
{y_m^{{\text{OSS}}} = {h_{Mm}}\sqrt {{P_M}} {x_M} + {n_m}},
\end{equation}
where subscript $m$ represents MU, ${h_{Mm}}$ denotes the fading gain of MBS-MU channel, and ${n_m}$ is the AWGN encountered at MU. Also, the received signal at SU is similarly written as
\begin{equation}\label{equa2}
{y_s^{{\text{OSS}}} = {h_{Ss}}\sqrt {{P_S}} {x_S} + {n_s},}
\end{equation}
where subscript $s$ represents SU, ${h_{Ss}}$ denotes the fading gain of SBS-SU channel, and ${n_s}$ is the AWGN encountered at SU. Meanwhile, due to the broadcast nature of wireless transmission, the signal transmissions of MBS and SBS may be overheard by E and the corresponding received signals are expressed as
\begin{equation}\label{equa3}
{y_{Me}^{{\text{OSS}}} = {h_{Me}}\sqrt {{P_M}} {x_M} + {n_e}}
\end{equation}
and
\begin{equation}\label{equa4}
{y_{Se}^{{\text{OSS}}} = {h_{Se}}\sqrt {{P_S}} {x_S} + {n_e},}
\end{equation}
where subscript $e$ represents the eavesdropper, ${h_{Me}}$ and ${h_{Se}}$ denote the fading gains of MBS-E and SBS-E channels, and ${n_e}$ is the AWGN encountered at E. Using the Shannon's capacity formula, we can obtain the channel capacity of MBS-MU and that of SBS-SU from (1) and (2) as
\begin{equation}\label{equa5}
{C_{Mm}^{{\text{OSS}}} = \alpha {\log _2}( {1 + {\gamma _M}|{h_{Mm}}{|^2}} ),}
\end{equation}
and
\begin{equation}\label{equa6}
{C_{Ss}^{{\text{OSS}}} = ( {1 - \alpha }){\log _2}( {1 + \gamma _S |{h_{Ss}}{|^2}}),}
\end{equation}
where $\alpha$ represents the fraction of the total spectrum assigned to the MBS-MU transmission, $\gamma _M  = {P_M}/{N_0}$ and $\gamma_S  = {P_S}/{N_0}$ are referred to as signal-to-noise ratios (SNRs) of MBS and SBS, respectively. Similarly, from (3) and (4), the channel capacity of MBS-E and that of SBS-E are obtained as
\begin{equation}\label{equa7}
{C_{Me}^{{\text{OSS}}} = \alpha {\log _2}( {1 + \gamma _M |{h_{Me}}{|^2}} ),}
\end{equation}
and
\begin{equation}\label{equa8}
{C_{Se}^{{\text{OSS}}} = ( {1 - \alpha } ){\log _2}( {1 + \gamma _S |{h_{Se}}{|^2}} ).}
\end{equation}
It is pointed out that $h_{Mm}$, $h_{Ss}$, $h_{Me}$ and $h_{Se}$ are modeled as independent zero-mean complex Gaussian random variables with respective variances of $\sigma^2_{Mm}$, $\sigma^2_{Ss}$, $\sigma^2_{Me}$ and $\sigma^2_{Se}$.

\subsection{Conventional IL-USS}
In this section, we present an interference-limited underlay spectrum sharing (IL-USS) scheme, where MBS and SBS are allowed to access the spectrum simultaneously and the transmit power of SBS shall be limited to ensure its induced interference to the MBS-MU transmission below a tolerable level for the QoS guarantee. As a consequence, when MBS sends its message ${x_M}$ to MU at a power of $P_M$, SBS simultaneously transmits ${x_S}$ to SU at a power of $P_s$ over the same spectrum band, leading to the fact that a mixed signal of ${x_M}$ and ${x_S}$ is received at MU and SU. Hence, we can express the received signals at MU and SU as
\begin{equation}\label{equa9}
{y_m^{{\text{IL-USS}}} = {h_{Mm}}\sqrt {{P_M}} {x_M} + {h_{Sm}}\sqrt {{P_S}} {x_S} + {n_m},}
\end{equation}
and
\begin{equation}\label{equa10}
{y_s^{{\text{IL-USS}}} = {h_{Ss}}\sqrt {{P_S}} {x_S} + {h_{Ms}}\sqrt {{P_M}} {x_M} + {n_s},}
\end{equation}
where ${h_{Sm}}$ and ${h_{Ms}}$ represent fading gains of SBS-MU and MBS-SU channels, respectively. Meanwhile, due to the broadcast nature of wireless transmissions, the eavesdropper may overhear the signal transmissions of MBS and SBS. Therefore, the corresponding received signal at E can be written as
\begin{equation}\label{equa11}
{y_e^{{\text{IL-USS}}} = {h_{Me}}\sqrt {{P_M}} {x_M} + {h_{Se}}\sqrt {{P_S}} {x_S} + {n_e}.}
\end{equation}
According to the Shannon's capacity formula and using (9) and (10), we may obtain the channel capacity of MBS-MU and that of SBS-SU as
\begin{equation}\label{equa12}
{C_{Mm}^{{\text{IL-USS}}} = {\log _2}( {1 + \frac{{{\gamma _M}|{h_{Mm}}{|^2}}}{{{\gamma _S}|{h_{Sm}}{|^2} + 1}}} ),}
\end{equation}
and
\begin{equation}\label{equa13}
{C_{Ss}^{{\text{IL-USS}}} = {\log _2}( {1 + \frac{{{\gamma _S}|{h_{Ss}}{|^2}}}{{{\gamma _M}|{h_{Ms}}{|^2} + 1}}} ).}
\end{equation}

Additionally, the eavesdropper attempts to interpret the confidential messages of $x_M$ and $x_S$ based on its received signal as given by (11). For simplicity, we here consider that the eavesdropper decodes $x_M$ and $x_S$ separately without using successive interference cancelation [36]. Indeed, more useful information may be tapped by an eavesdropper with the successive interference cancelation, which can be similarly adopted at the MU and SU for achieving a better transmission reliability performance. Hence, no additional benefits can be achieved for our heterogeneous cellular networks from an SRT perspective by employing the successive interference cancelation at both the MU and SU as well as the eavesdropper. It is of interest to explore an impact of successive interference cancelation on the secrecy performance of our IC-USS scheme, which may be considered for future work. Treating $x_S$ as interference, we obtain the channel capacity of MBS-E from (11) as
\begin{equation}\label{equa14}
{C_{Me}^{{\text{IL-USS}}} = {\log _2}( {1 + \frac{{{\gamma _M}|{h_{Me}}{|^2}}}{{{\gamma _S}|{h_{Se}}{|^2} + 1}}}).}
\end{equation}
Also, the eavesdropper may decode $x_s$ based on (11) and the channel capacity of SBS-E is similarly given by
\begin{equation}\label{equa15}
{C_{Se}^{{\text{IL-USS}}} = {\log _2}( {1 + \frac{{{\gamma _S}|{h_{Se}}{|^2}}}{{{\gamma _M}|{h_{Me}}{|^2} + 1}}} ).}
\end{equation}

\subsection{Proposed IC-USS}
{{In this section, we propose an interference-canceled underlay spectrum sharing (IC-USS) scheme, where MBS and SBS simultaneously transmit their signals of $x_M$ and $x_S$ over the same spectrum band and may interfere with each other. In order to cancel out the interference received at MU from SBS, a special signal denoted by $x_m$ is sophisticatedly designed and transmitted at MBS. This means that MBS shall transmit a mixed signal of $x_M$ and $x_m$ simultaneously, which are utilized to carry the desired information and to cancel out the interference received at MU, respectively. For notational convenience, let $P_m$ and $\bar P_m$ denote the instantaneous and average transmit power of $x_m$, respectively. Moreover, the transmit power of $x_M$ is considered to be $P_M - \bar P_m$, where $P_M \ge 0$ and $0 \le \bar P_m \le P_M$. This implies that an average transmit power of the mixed signal of $x_M$ and $x_m$ is $P_M$, which guarantees a fair comparison with the IL-USS scheme in terms of the average power consumption. While MBS sends its mixed signal of $x_M$ and $x_m$, SBS also transmits $x_S$ with a weight coefficient $w_S$ at a power of $P_S$, wherein $E(|w_S|^2 = 1)$. Thus, we can express the received signal at MU as}}
\begin{equation}\label{equa16}
\begin{split}
  y_m^{{\text{IC-USS}}} &= {h_{Mm}}( {\sqrt {{P_M} - {\bar P_m}} {x_M} + {x_m}} ) \\
  &\quad + {h_{Sm}}\sqrt {{P_S}} w_S {x_S} + {n_m} \\
   &= {h_{Mm}}\sqrt {{P_M} - {\bar P_m}} {x_M} \\
   &\quad + (  {{h_{Mm}}{x_m} + \sqrt {{P_S}} {h_{Sm}}w_S{x_S}} ) + {n_m}.  \\
\end{split}
\end{equation}
{{In order to cancel out the interference received at MU, both the specially-designed signal ${x_m}$ and the weight coefficient $w_S$ should satisfy the following equality}}
\begin{equation}\label{equa17}
{h_{Mm}}{x_m} + \sqrt {{P_S}} {h_{Sm}}w_S {x_S} = 0,
\end{equation}
{{from which infinite solutions are available for the interference neutralization. Throughout this paper, a solution of $[x_m, w_S]$ to the preceding equation is given by}}
\begin{equation}\label{equa18}
[x_m, w_S] = \frac{1}{{{\sigma _{Mm}}}}[- {\sqrt {{P_s}} }|{h_{Sm}}|{e^{ - j{\theta _{Mm}}}}{x_S},|{h_{Mm}}|{e^{ - j{\theta _{Sm}}}}],
\end{equation}
{{where $\sigma _{Mm}^2 = E(|{h_{Mm}}{|^2})$ is the channel variance of MBS-MU, ${{\theta _{Mm}}}$ and ${{\theta _{Sm}}}$ denote the channel phase of MBS-MU and that of SBS-MU, respectively. It can be observed from (18) that ${h_{Mm}}$, ${h_{Sm}}$, $\sigma^2_{Mm}$, $P_S$ and ${x_S}$ should be known at MBS and SBS for an appropriate design of $[x_m, w_S]$. Thanks to that MBS and SBS are connected the core network through wire cables as shown in Fig. 1, MBS can easily acquire the exact information of $P_S$ and ${x_S}$ from SBS via the core network. Moreover, the CSIs of ${h_{Mm}}$, ${h_{Sm}}$ and $\sigma^2_{Mm}$ could be estimated at MU and then fed back to MBS and SBS [37]. From (18), one can readily obtain the instantaneous and average transmit powers of $x_m$ as}}
\begin{equation}\label{equa19}
[{P_m}, \bar P_m] = [ \frac{{|{h_{Sm}}{|^2}}}{{\sigma _{Mm}^2}}{P_S} ,\frac{{\sigma _{Sm}^2}}{{\sigma _{Mm}^2}}{P_S}],
\end{equation}
where $\sigma _{Sm}^2 = E(|{h_{Sm}}{|^2})$ is the channel variance of SBS-MU. Noting that the average transmit power of $x_m$ (i.e., $\bar P_m$) should be in the range of $0 \le \bar P_m \le P_M$ and using (19), we obtain the following inequality
\begin{equation}\label{equa20}
\frac{{{P_M}}}{{{P_S}}} \ge \frac{{\sigma _{Sm}^2}}{{\sigma _{Mm}^2}},
\end{equation}
which is a necessary condition for MBS to cancel out the interference received at MU. In other words, as long as the transmit power $P_M$ of MBS is sufficiently high to satisfy (20), it is possible to perfectly cancel out the interference received at MU from SBS by employing a specially-designed signal $x_m$ and weigh coefficient $w_S$ of (18). It is worth mentioning that although the sophisticatedly-designed signal $x_m$ can be employed to neutralize the mutual interference as well as to confuse the eavesdropper, it comes at the cost of consuming partial transmit power that could be used for transmitting the desired information-bearing signal $x_M$. It is of interest to investigate the power allocation between the sophisticatedly-designed signal and information-bearing signal in terms of maximizing the secrecy rate of wireless transmissions. Moreover, an optimal solution of $[x_m, w_S] $ may also be considered for further enhancing the secrecy performance, which is beyond the scope of this paper. Substituting $x_m$ and $w_S$ from (18) into (16) gives
\begin{equation}\label{equa21}
{y_m^{{\text{IC-USS}}} = {h_{Mm}}\sqrt {{P_M} - {\bar P_m}} {x_M} + {n_m}},
\end{equation}
where $\bar P_m$ is given by (19). Also, we can express the received signal at SU as
\begin{equation}\label{equa22}
y_s^{{\text{IC-USS}}} = {h_{Ss}}\sqrt {{P_S}} w_S {x_S} + {h_{Ms}}( {\sqrt {{P_M} - {\bar P_m}} {x_M} + {x_m}} ) + {n_s},
\end{equation}
where $x_m$ and $w_S$ are given by (18). Meanwhile, the eavesdropper may overhear the signal transmissions from MBS and SBS, thus the corresponding received signal can be written as
\begin{equation}\label{equa23}
y_e^{{\text{IC-USS}}} = {h_{Me}}( {\sqrt {{P_M} - {\bar P_m}} {x_M} + {x_m}} )  + {h_{Se}}\sqrt {{P_S}} {w_S x_S} + {n_e}.
\end{equation}
Applying the Shannon's capacity formula to (21), we can obtain the channel capacity of MBS-MU transmission relying on the proposed IC-USS scheme as
\begin{equation}\label{equa24}
C_{Mm}^{{\text{IC-USS}}} = {\log _2}[1 + ({\gamma _M} - {{\bar \gamma }_m})|{h_{Mm}}{|^2}],
\end{equation}
where ${{\gamma }_M} = {{P}_M}/{N_0}$ and ${{\bar \gamma }_m} = {{\bar P}_m}/{N_0}$. Similarly, by treating both $x_M$ and $x_m$ as interference, the channel capacity of SBS-SU transmission relying on our IC-USS scheme can be obtained from (22) as
\begin{equation}\label{equa25}
C_{Ss}^{{\text{IC-USS}}} = {\log _2}[1 + \frac{{|{h_{Ss}}{|^2}{{\left| {{h_{Mm}}} \right|}^2}{\gamma _S}/\sigma _{Mm}^2}}{{|{h_{Ms}}{|^2}({\gamma _M} - {{\bar \gamma }_m} + {\gamma _m}) + 1}}],
\end{equation}
where ${{\gamma }_S} = {{P}_S}/{N_0}$ and ${{\gamma }_m} = {{P}_m}/{N_0}$.
Moreover, the eavesdropper may exploit its overheard signal of (23) to decode the confidential messages of $x_M$ and $x_S$. Similar to the conventional IL-USS, we consider that $x_M$ and $x_S$ are decoded independently at the eavesdropper without using successive interference cancelation. Thus, treating $x_m$ and $x_S$ as interference, we obtain the MBS-E channel capacity from (23) as
\begin{equation}\label{equa26}
C_{Me}^{{\text{IC-USS}}} = {\log _2}[1 + \frac{{|{h_{Me}}{|^2}({\gamma _M} - {{\bar \gamma }_m})}}{{|{h_{Me}}{|^2}{\gamma _m} + |{h_{Se}}{|^2}|{h_{Mm}}{|^2}{\gamma _S}/\sigma _{Mm}^2 + 1}}].
\end{equation}
Similarly, by using (23) and treating $x_M$ and $x_m$ as interference, the SBS-E channel capacity can be given by
\begin{equation}\label{equa27}
C_{Se}^{{\text{IC-USS}}} = {\log _2}[1 + \frac{{|{h_{Se}}{|^2}{{\left| {{h_{Mm}}} \right|}^2}{\gamma _S}/\sigma _{Mm}^2}}{{|{h_{Me}}{|^2}({\gamma _M} - {{\bar \gamma }_m} + {\gamma _m}) + 1}}],
\end{equation}
which completes the system model of our IC-USS scheme.

\section{SRT Analysis of Spectrum Sharing Schemes}
In this section, we present the SRT analysis of conventional OSS and IL-USS as well as proposed IC-USS schemes over Rayleigh fading channels. As discussed in [38], the security and reliability of wireless communications are characterized by the intercept probability and outage probability experienced at the eavesdropper and legitimate receiver, respectively. Let us first recall the definitions of intercept probability and outage probability. According to physical-layer security literature [38], [39], a source message with a secrecy rate of $R_s$ needs to be encoded by a secrecy encoder, generating an overall codeword with an increased rate $R_o$ to be transmitted to the destination. It is pointed out that the rate difference $R_o - R_s$ represents an extra redundancy introduced for the sake of defending against eavesdropping. The definition of intercept probability and outage probability is detailed as follows.\\
\textbf{Definition 1}: \emph{According to the Shannon's coding theorem, when the capacity of the main channel spanning from the source to legitimate destination falls below the transmission rate {{$R_o$}}, it is impossible for the destination to successfully decode the source message and an outage event occurs in this case. Thus, by letting $C_m$ denote the capacity of main channel, the probability of occurrence of outage event (referred to as outage probability) is expressed as}
\begin{equation}\label{equa28}
{{P_{\textrm{out}}} = \Pr \left( {{C_m} < {R_o}} \right)}.
\end{equation}
\textbf{Definition 2}:
\emph{As discussed in [38] and [39], if the capacity of a wiretap channel is higher than the rate difference of {{$R_o-R_s$}}, perfect secrecy is not achievable and an intercept event is considered to happen. Hence, the probability of occurrence of intercept event (called intercept probability) is given by}
\begin{equation}\label{equa29}
{{P_{\textrm{int}}} = \Pr \left( {{C_e} > {R_o} - {R_s}} \right),}
\end{equation}
\emph{where $C_e$ represents the capacity of wiretap channel.}

One can observe from (28) and (29) that the outage probability and intercept probability affect each other with respect to an intermediate parameter $R_o$. In what follows, we present the analysis of outage probability and intercept probability for the OSS, IL-USS and IC-USS schemes for the sake of quantitatively characterizing their security and reliability relationship.

\subsection{Conventional OSS}
This subsection analyzes the outage probability and intercept probability of macro-cell and small-cell transmissions relying on the conventional OSS approach. Without loss of any generality, let $R_M^o$ and $R_S^o$ represent the overall data rates of MBS-MU and SBS-SU transmissions, respectively. From (28), the outage probability of MBS-MU transmission is obtained as
\begin{equation}\label{equa30}
{P_{Mm{\text{-out}}}^{{\text{OSS}}} = \Pr \left( {C_{Mm}^{{\text{OSS}}} < R_M^o} \right) },
\end{equation}
where $C_{Mm}^{{\text{OSS}}}$ is given by (5). Substituting $C_{Mm}^{{\text{OSS}}}$ from (5) into (30) yields
\begin{equation}\label{equa31}
\begin{split}
P_{Mm{\text{-out}}}^{{\text{OSS}}} &= \Pr \left( {\alpha {{\log }_2}( {1 + {\gamma _M}|{h_{Mm}}{|^2}} ) < R_M^o} \right)  \\
&= \Pr \left( { {|{h_{Mm}}{|^2} < \Delta_M} } \right),
\end{split}
\end{equation}
where $\Delta_M = ({{2^{\frac{{R_M^o}}{\alpha }}} - 1})/{{{\gamma _M}}}$. Noting that $|{h_{Mm}}{|^2}$ is an exponentially distributed random variable with a mean of $\sigma _{Mm}^2$, we arrive at
\begin{equation}\label{equa32}
{P_{Mm{\text{-out}}}^{{\text{OSS}}} = 1 - \exp ( - \frac{\Delta_M }{{\sigma _{Mm}^2}})}.
\end{equation}
Similarly, using (6) and (28), the outage probability of SBS-SU transmission is given by
\begin{equation}\label{equa33}
{P_{Ss{\text{-out}}}^{{\text{OSS}}} = \Pr \left( {C_{Ss}^{{\text{OSS}}} < R_S^o} \right). }
\end{equation}
Substituting $C_{Ss}^{\text{OSS}}$ from (6) into (33) yields
\begin{equation}\label{equa34}
P_{Ss{\text{-out}}}^{{\text{OSS}}} = 1 - \exp ( - \frac{\Delta_S }{{\sigma _{Ss}^2}}),
\end{equation}
where $\Delta_S = ({2^{\frac{{R_S^o}}{{1 - \alpha }}}} - 1)/\gamma_S$ and $\sigma _{Ss}^2$ is an expected value of the exponentially distributed random variable of $|{h_{Ss}}{|^2}$.

Additionally, for notational convenience, let $R_M^s$ and $R_S^s$ denote secrecy rates of MBS-MU and SBS-SU transmissions, respectively. Moreover, the random variables of $|{h_{Me}}{|^2}$ and $|{h_{Se}}{|^2}$ are independent exponentially distributed with respective means of $\sigma _{Me}^2$ and $\sigma _{Se}^2$. Using (7) and (29), we can obtain the intercept probability of MBS-E wiretap channel as
\begin{equation}\label{equa35}
P_{Me{\text{-int}}}^{{\text{OSS}}} = \Pr \left( {C_{Me}^{{\text{OSS}}} > R_M^o - R_M^s} \right){\kern 1pt}  =\exp ( - \frac{\Delta^{d}_{M}}{{\sigma _{Me}^2}}),
\end{equation}
where $\Delta^{d}_{M} = ({{2^{\frac{{R_M^o - R_M^s}}{\alpha }}} - 1})/{{{\gamma _M}}}$. Similarly, from (8) and (29), the intercept probability of SBS-E wiretap channel is given by
\begin{equation}\label{equa36}
{P_{Se{\text{-int}}}^{{\text{OSS}}} = \Pr \left( {C_{Se}^{{\text{OSS}}} > R_S^o - R_S^s} \right) =\exp ( - \frac{\Delta^{d}_{S}}{{\sigma _{Se}^2}}),}
\end{equation}
where $\Delta^{d}_{S} = ({{2^{\frac{{R_S^o - R_S^s}}{1-\alpha }}} - 1})/{{{\gamma _S}}}$.

So far, we have derived closed-form expressions of outage probability and intercept probability for macro-cell and small-cell transmissions separately, as shown in (32) and (34)-(36). In order to show a coupled effect between the macro-cell and small-cell transmissions, we here define an overall outage probability of the macro cell and small cell by the product of their individual outage probabilities. As a consequence, an overall outage probability for the conventional OSS scheme can be expressed as
\begin{equation}\label{equa37}
P_{{\text{out}}}^{{\text{OSS}}} = P_{Mm{\text{-out}}}^{{\text{OSS}}} \times P_{Ss{\text{-out}}}^{{\text{OSS}}} ,
\end{equation}
where $P_{Mm{\text{-out}}}^{{\text{OSS}}}$ and $P_{Ss{\text{-out}}}^{{\text{OSS}}}$ are given by (32) and (34), respectively. Similarly, an overall intercept probability of the heterogeneous macro cell and small cell can be defined as the product of their individual intercept probabilities. Hence, an overall intercept probability for the conventional OSS scheme is written as
\begin{equation}\label{equa38}
P_{{\text{int}}}^{{\text{OSS}}} = P_{Me{\text{-int}}}^{{\text{OSS}}} \times P_{Se{\text{-int}}}^{{\text{OSS}}} ,
\end{equation}
where $P_{Me{\text{-int}}}^{{\text{OSS}}}$ and $P_{Se{\text{-int}}}^{{\text{OSS}}}$ are given by (35) and (36), respectively.

\subsection{Conventional IL-USS}
This subsection presents the SRT analysis of IL-USS scheme. Using (12) and (28) and noting that $|{h_{Mm}}{|^2}$ and $|{h_{Sm}}{|^2}$ are independent exponentially distributed random variables with respective means of $\sigma _{Mm}^2$ and $\sigma _{Sm}^2$, we can obtain the outage probability of MBS-MU transmission relying on the IL-USS scheme as
\begin{equation}\label{equa39}
\begin{split}
  P_{Mm{\text{-out}}}^{{\text{IL-USS}}} &= \Pr \left( {C_{Mm}^{{\text{IL-USS}}} < R_M^o} \right)  \\
&= \Pr \left( |{h_{Mm}}{|^2} < ({\gamma _S}|{h_{Sm}}{|^2} + 1) \Lambda _M \right)  \\
&= 1 - \frac{{\sigma _{Mm}^2}}{{{\gamma _S}\sigma _{Sm}^2{\Lambda _M} + \sigma _{Mm}^2}}\exp ( { - \frac{{{\Lambda _M}}}{{\sigma _{Mm}^2}}} ),  \\
\end{split}
\end{equation}
where ${\Lambda _M} = ( {{2^{R_M^o}} - 1} )/{\gamma _M}$. From (13) and (28), the outage probability of SBS-SU transmission is given by
\begin{equation}\label{equa40}
\begin{split}
  P_{Ss{\text{-out}}}^{{\text{IL-USS}}} & = \Pr \left( {C_{Ss}^{{\text{IL-USS}}} < R_S^o} \right)\\
  &  = \Pr \left( |{h_{SS}}{|^2} < ({\gamma _M}|{h_{Ms}}{|^2} + 1)\Lambda _S \right)  \\
  & = 1 - \frac{{\sigma _{Ss}^2}}{{{\gamma _M}{\sigma_{Ms}^2\Lambda _S} + \sigma _{Ss}^2}}\exp ( { - \frac{{{\Lambda _S}}}{{\sigma _{Ss}^2}}} ), \\
\end{split}
\end{equation}
where ${\Lambda _S} = ( {{2^{R_S^o}} - 1} )/{\gamma _S}$ and $\sigma _{Ms}^2$ is a mean of the exponentially distributed random variable $|{h_{Ms}}{|^2}$. Similar to (37), an overall outage probability for the conventional IL-USS scheme can be obtained as
\begin{equation}\label{equa41}
P_{{\text{out}}}^{{\text{IL-USS}}} = P_{Mm{\text{-out}}}^{{\text{IL-USS}}} \times P_{Ss{\text{-out}}}^{{\text{IL-USS}}},
\end{equation}
where $P_{Mm{\text{-out}}}^{{\text{IL-USS}}}$ and $P_{Ss{\text{-out}}}^{{\text{IL-USS}}}$ are given by (39) and (40), respectively.

Additionally, from (14), (15) and (29), we can obtain intercept probabilities experienced over MBS-E and SBS-E wiretap channels as
\begin{equation}\label{equa42}
\begin{split}
  P_{Me{\text{-int}}}^{{\text{IL-USS}}} &= \Pr \left( {C_{Me}^{{\text{IL-USS}}} > R_M^o - R_M^s} \right)  \\
&= \Pr \left( |{h_{Me}}{|^2} > ({\gamma _S}|{h_{Se}}{|^2} + 1) \Lambda _M^d  \right)  \\
& = \frac{{\sigma _{Me}^2}}{{\sigma _{Me}^2 + {\gamma _S}\sigma _{Se}^2\Lambda _M^d}}\exp ( { - \frac{{\Lambda _M^d}}{{\sigma _{Me}^2}}} ),  \\
\end{split}
\end{equation}
and
\begin{equation}\label{equa43}
\begin{split}
  P_{Se{\text{-int}}}^{{\text{IL-USS}}} &= \Pr \left( {C_{Se}^{{\text{IL-USS}}} > R_S^o - R_S^s} \right)  \\
&= \Pr \left( |{h_{Se}}{|^2} > ({\gamma _M}|{h_{Me}}{|^2} + 1) \Lambda _S^d  \right)  \\
&= \frac{{\sigma _{Se}^2}}{{\sigma _{Se}^2 + {\gamma _M}\sigma _{Me}^2\Lambda _S^d}}\exp ( { - \frac{{\Lambda _S^d}}{{\sigma _{Se}^2}}} ),  \\
\end{split}
\end{equation}
where $\Lambda _M^d = ( {{2^{R_M^o - R_M^s}} - 1} )/{\gamma _M}$ and $\Lambda _S^d = ( {{2^{R_S^o - R_S^s}} - 1} )/{\gamma _S}$. Therefore, an overall intercept probability of the conventional IL-USS scheme can be similarly defined as the product of individual intercept probabilities of $P_{Me{\text{-int}}}^{{\text{IL-USS}}}$ and $P_{Se{\text{-int}}}^{{\text{IL-USS}}}$, namely
\begin{equation}\label{equa44}
P_{{\text{int}}}^{{\text{IL-USS}}} = P_{Me{\text{-int}}}^{{\text{IL-USS}}} \times P_{Se{\text{-int}}}^{{\text{IL-USS}}},
\end{equation}
where $P_{Me{\text{-int}}}^{{\text{IL-USS}}}$ and $P_{Se{\text{-int}}}^{{\text{IL-USS}}}$ are given by (42) and (43), respectively.

\subsection{Proposed IC-USS}
In this subsection, we carry out the SRT analysis of proposed IC-USS scheme by deriving its closed-form outage probability and intercept probability. From (24) and (28), an individual outage probability of the MBS-MU transmission relying on the proposed IC-USS scheme is given by
\begin{equation}\label{equa45}
\begin{split}
P_{Mm{\text{-out}}}^{{\text{IC-USS}}} &= \Pr \left( {C_{Mm}^{{\text{IC-USS}}} < R_M^o} \right) \\
&= 1 - \exp ( { - \frac{{{\Lambda _M}}}{{\sigma _{Mm}^2 - \beta\sigma _{Sm}^2}}}),
\end{split}
\end{equation}
where $\beta = \gamma_S/\gamma_M$ is referred to as the small-to-macro ratio (SMR). Using (25), we can obtain an individual outage probability of the SBS-SU transmission for the IC-USS scheme as
\begin{equation}\nonumber
\begin{split}
&P_{Ss{\text{-out}}}^{{\text{IC-USS}}} = \Pr \left( {\frac{{|{h_{Ss}}{|^2}{{\left| {{h_{Mm}}} \right|}^2}}}{{|{h_{Ms}}{|^2}({\gamma _M} - {{\bar \gamma }_m} + {\gamma _m}) + 1}} < {\Lambda _S}\sigma _{Mm}^2} \right)\\
&=\Pr \left( {\frac{{|{h_{Ss}}{|^2}{{\left| {{h_{Mm}}} \right|}^2}}}{{|{h_{Ms}}{|^2}({\Lambda _S}\sigma _{Mm}^2{\gamma _M} + {X_{Sm}}) + {\Lambda _S}\sigma _{Mm}^2}} < 1} \right)
\end{split}
\end{equation}
where ${X_{Sm}} = ({2^{R_S^o}} - 1)(|{h_{Sm}}{|^2} - \sigma _{Sm}^2)$. It is challenging to derive an exact closed-form expression for $P_{Ss{\text{-out}}}^{{\text{IC-USS}}}$. Let us consider an asymptotic case of ${2^{R_S^o}} \sigma _{Sm}^2 \to 0$, for which the random variable ${X_{Sm}} $ is equal to zero with the probability of one, since its mean and variance both approach to zero. Thus, we have
\begin{equation}\label{equa46}
\begin{split}
&P_{Ss{\text{-out}}}^{{\text{IC-USS}}} = \Pr \left( {\frac{{|{h_{Ss}}{|^2}{{\left| {{h_{Mm}}} \right|}^2}}}{{|{h_{Ms}}{|^2}{\Lambda _S}\sigma _{Mm}^2{\gamma _M} + {\Lambda _S}\sigma _{Mm}^2}} < 1} \right)\\
& = 1 - \int_0^\infty  {\frac{{\sigma _{Ss}^2x}}{{\sigma _{Ms}^2{\Lambda _S}{\gamma _M} + \sigma _{Ss}^2x}}\exp ( - x - \frac{{{\Lambda _S}}}{{\sigma _{Ss}^2x}})dx},
 \end{split}
\end{equation}
for ${2^{R_S^o}} \sigma _{Sm}^2 \to 0$. Using (45) and (46), we can obtain an overall outage probability for the proposed IC-USS scheme as
\begin{equation}\label{equa47}
P_{{\text{out}}}^{{\text{IC-USS}}} = P_{Mm{\text{-out}}}^{{\text{IC-USS}}} \times P_{Ss{\text{-out}}}^{{\text{IC-USS}}}.
\end{equation}

In addition, from (26) and (29), an intercept probability encountered by MBS-E wiretap channel relying on the proposed IC-USS scheme is given by
\begin{equation}\label{equa48}
\begin{split}
&P_{Me{\text{-int}}}^{{\text{IC-USS}}} = \Pr \left( {C_{Me}^{{\text{IC-USS}}} > R_M^o - R_M^s} \right) \\
&= \Pr \left( {\frac{{|{h_{Me}}{|^2}({\gamma _M} - {{\bar \gamma }_m})}}{{|{h_{Me}}{|^2}{\gamma _m} + |{h_{Se}}{|^2}{X_{Mm}}{\gamma _S} + 1}} > {\gamma _M}\Lambda _M^d} \right) \\
&= \Pr \left( {|{h_{Me}}{|^2}{Y_{Sm}} > \Lambda _M^d(|{h_{Se}}{|^2}{X_{Mm}}{\gamma _S} + 1)} \right), \\
\end{split}
\end{equation}
where ${X_{Mm}} = \frac{{|{h_{Mm}}{|^2}}}{{\sigma _{Mm}^2}}$ and ${Y_{Sm}} = 1 - \frac{\beta }{{\sigma _{Mm}^2}}[\sigma _{Sm}^2 + |{h_{Sm}}{|^2}({2^{R_M^o - R_M^s}} - 1)]$. We can rewrite (48) as
\begin{equation}\nonumber
\begin{split}
&P_{Me{\textrm{-}}{\mathop{\textrm{int}}} }^{{\textrm{IC-USS}}} = \Pr ({Y_{Sm}} > 0)\Pr ( {|{h_{Me}}{|^2}{Y_{Sm}}} \\
&\quad\quad\quad\quad\quad\quad\quad > \Lambda _M^d(|{h_{Se}}{|^2}{X_{Mm}}{\gamma _S} + 1)|{Y_{Sm}} > 0 ).
\end{split}
\end{equation}
Here, we also consider an asymptotic case of ${2^{R_M^o}} \sigma _{Sm}^2 \to 0$, for which an equality $|{h_{Sm}}{|^2}({{2^{R_M^o-R^s_M}-1}}) = \sigma _{Sm}^2({2^{R_M^o-R^s_M}-1})$ holds with the probability of one, leading to ${Y_{Sm}} = 1 - \sigma _{Sm}^2\beta {2^{R_M^o - R_M^s}}/\sigma _{Mm}^2$ and $\Pr ({Y_{Sm}} > 0) = 1$. Substituting these results into the preceding equation yields
\begin{equation}\nonumber
P_{Me{\textrm{-}}{\mathop{\textrm{int}}} }^{{\textrm{IC-USS}}} = \Pr \left( {|{h_{Me}}{|^2}\Omega  > \Lambda _M^d(|{h_{Se}}{|^2}{X_{Mm}}{\gamma _S} + 1)} \right) ,
\end{equation}
where $\Omega = 1 - \sigma _{Sm}^2\beta {2^{R_M^o - R_M^s}}/\sigma _{Mm}^2$. Noting that $|h_{Me}|^2$, $|h_{Se}|^2$ and $X_{Mm}$ are independent exponentially distributed random variables with respective means of $\sigma^2_{Me}$, $\sigma^2_{Se}$ and $1$, we arrive at
\begin{equation}\label{equa49}
P_{Me{\textrm{-}}{\mathop{\textrm{int}}} }^{{\textrm{IC-USS}}} = \Theta \exp (\Theta  - \frac{{\Lambda _M^d}}{{\sigma _{Me}^2\Omega }})Ei(\Theta ),
\end{equation}
for ${2^{R_M^o}} \sigma _{Sm}^2 \to 0$, where $\Theta  = \frac{{\Omega \sigma _{Me}^2}}{{\Lambda _M^d\sigma _{Se}^2{\gamma _s}}}$ and $Ei(\Theta ) = \int_\Theta ^\infty  {{t^{ - 1}}{e^{ - t}}dt} $. Using (27) and (29), we can obtain an intercept probability experienced over SBS-E wiretap channel for the proposed IC-USS scheme as
\begin{equation}\label{equa50}
\begin{split}
&  P_{Se{\text{-int}}}^{{\text{IC-USS}}}  = \Pr \left( {C_{Se}^{{\text{IC-USS}}} > R_S^o - R_S^s} \right)  \\
&= \Pr \left( {\frac{{|{h_{Se}}{|^2}{{\left| {{h_{Mm}}} \right|}^2}}}{{|{h_{Me}}{|^2}(\Lambda _S^d\sigma _{Mm}^2{\gamma _M} + {Z_{Sm}}) + \Lambda _S^d\sigma _{Mm}^2}} > 1} \right),
\end{split}
\end{equation}
where ${Z_{Sm}} = ({2^{R_S^o - R_S^s}} - 1)(|{h_{Sm}}{|^2} - \sigma _{Sm}^2)$. Considering ${2^{R_S^o}} \sigma _{Sm}^2 \to 0$, one can conclude that ${Z_{Sm}} $ approaches to zero with the probability of one. Hence, we have
\begin{equation}\label{equa51}
\begin{split}
&P_{Se{\text{-int}}}^{{\text{IC-USS}}}  = \Pr \left( {\frac{{|{h_{Se}}{|^2}{{\left| {{h_{Mm}}} \right|}^2}}}{{|{h_{Me}}{|^2}\Lambda _S^d\sigma _{Mm}^2{\gamma _M} + \Lambda _S^d\sigma _{Mm}^2}} > 1} \right) \\
&= \int_0^\infty  {\frac{{\sigma _{Se}^2x}}{{\sigma _{Me}^2\Lambda _S^d{\gamma _M} + \sigma _{Se}^2x}}\exp ( - x - \frac{{\Lambda _S^d}}{{\sigma _{Se}^2x}})dx},  \\
\end{split}
\end{equation}
for ${2^{R_S^o}} \sigma _{Sm}^2 \to 0$. Similar (38), an overall intercept probability for the proposed IC-USS scheme is obtained as
\begin{equation}\label{equa52}
P_{{\text{int}}}^{{\text{IC-USS}}} = P_{Me{\text{-int}}}^{{\text{IC-USS}}} \times P_{Se{\text{-int}}}^{{\text{IC-USS}}},
\end{equation}
where $P_{Me{\text{-int}}}^{{\text{IC-USS}}}$ and $P_{Se{\text{-int}}}^{{\text{IC-USS}}}$ are given by (49) and (51), respectively.

\section{Secrecy Diversity Gain Analysis}
In this section, we present the secrecy diversity gain analysis for the conventional OSS and IL-USS as well as the proposed IC-USS schemes by characterizing an asymptotic behavior of the outage probability with an intercept probability constraint in the high SNR region. To be specific, we first analyze an asymptotic outage probability as a function of the intercept probability with $\gamma_M  \to \infty $, and then derive the secrecy diversity gain as a ratio of the logarithmic asymptotic outage probability to the logarithmic SNR ${\gamma_M}$, as mathematically described below
\begin{equation}\label{equa53}
{{d_s} =  - \mathop {\lim }\limits_{{\gamma_M} \to \infty } \frac{{\log {P_{{\text{out}}}}\left( {{\gamma_M, P_{{\text{int}}}}} \right)}}{{\log {\gamma_M}}},}
\end{equation}
where ${P_{{\text{out}}}}\left( {{\gamma_M, P_{{\text{int}}}}} \right)$ represents an outage probability as a function of SNR $\gamma_M$ and an intercept probability constraint $P_{{\text{int}}}$.

\subsection{Conventional OSS}
This subsection analyzes the secrecy diversity gain of conventional OSS scheme, which is described as
\begin{equation}\label{equa54}
{d_s^{\textrm{OSS}} =  - \mathop {\lim }\limits_{\gamma_M  \to \infty } \frac{{\log \left( {P_{{\text{out}}}^{{\text{OSS}}}} \right)}}{{\log \gamma_M }}},
\end{equation}
where $P_{{\text{out}}}^{{\text{OSS}}}$ is the outage probability of OSS scheme as given by (37). Substituting $P_{{\text{out}}}^{{\text{OSS}}}$ from (37) into (54) gives
\begin{equation}\label{equa55}
{d_s^{\textrm{OSS}} =  - \mathop {\lim }\limits_{\gamma_M  \to \infty } \frac{{\log \left( {P_{Mm{\text{-out}}}^{{\text{OSS}}} \cdot P_{Ss{\text{-out}}}^{{\text{OSS}}}} \right)}}{{\log \gamma_M }}.}
\end{equation}
From (35) and (36), we can have
\begin{equation}\label{equa56}
{{2^{^{\frac{{R_M^o}}{\alpha }}}} = {2^{\frac{{R_M^s}}{\alpha }}}\left( {1 - \gamma_M \sigma _{Me}^2\ln P_{Me{\text{-int}}}^{{\text{OSS}}}} \right),}
\end{equation}
and
\begin{equation}\label{equa57}
{{2^{^{\frac{{R_S^o}}{{1 - \alpha }}}}} = {2^{\frac{{R_S^s}}{{1 - \alpha }}}}\left( {1 - {\gamma _S}\sigma _{Se}^2\ln P_{Se{\text{-int}}}^{{\text{OSS}}}} \right).}
\end{equation}
Substituting ${2^{^{\frac{{R_M^o}}{\alpha }}}}$ and ${2^{^{\frac{{R_S^o}}{{1 - \alpha }}}}}$ from (56) and (57), respectively, into (32) and (34) yields
\begin{equation}\label{equa58}
{P_{Mm{\text{-out}}}^{{\text{OSS}}} = 1 - \exp \left( { - \frac{{{2^{\frac{{R_M^s}}{\alpha }}}\left( {1 - \gamma_M \sigma _{Me}^2\ln P_{Me{\text{-int}}}^{{\text{OSS}}}} \right) - 1}}{{\gamma_M \sigma _{Mm}^2}}} \right),}
\end{equation}
and
\begin{equation}\label{equa59}
{P_{Ss{\text{-out}}}^{{\text{OSS}}} = 1 - \exp \left( { - \frac{{{2^{\frac{{R_S^s}}{{1 - \alpha }}}}\left( {1 - {\gamma _S}\sigma _{Se}^2\ln P_{Se{\text{-int}}}^{{\text{OSS}}}} \right) - 1}}{{{\gamma _S}\sigma _{Ss}^2}}} \right).}
\end{equation}
From (58), one can readily obtain
\begin{equation}\label{equa60}
\mathop {\lim }\limits_{{\gamma _M} \to \infty } P_{Mm\textrm{-out}}^{\textrm{OSS}} = 1 - {(P_{Me\textrm{-int}}^{\textrm{OSS}})^{{{2^{(R_M^s/\alpha )}}\sigma _{Me}^2/\sigma _{Mm}^2}}},
\end{equation}
where $P_{Me\textrm{-int}}^{\textrm{OSS}}$ is an intercept probability required for macro-cell transmissions. Similarly, denoting $\gamma_S = \beta \gamma_M$ wherein $0 \le \beta \le \frac{{\sigma _{Mm}^2}}{{\sigma _{Sm}^2}}$ as implied from (20), we have
\begin{equation}\label{equa61}
\mathop {\lim }\limits_{{\gamma _M} \to \infty } P_{Ss\textrm{-out}}^{\textrm{OSS}} = 1 - {(P_{Se\textrm{-int}}^{\textrm{OSS}})^{{2^{R_S^s/(1 - \alpha )}}\sigma _{Se}^2/\sigma _{Ss}^2}},
\end{equation}
where $P_{Se\textrm{-int}}^{\textrm{OSS}}$ is a required intercept probability for small-cell transmissions. One can readily observe from (60) and (61) that the outage probabilities for both macro-cell and small-cell transmissions are decreasing functions with regard to the intercept probability, showing a tradeoff between the security and reliability. Moreover, as the required intercept probability decreases to zero, the corresponding outage probability increases to one, and vice versa. Substituting $P_{Mm\textrm{-out}}^{\textrm{OSS}} $ and $P_{Ss\textrm{-out}}^{\textrm{OSS}} $ from (60) and (61) into (55) gives
\begin{equation}\label{equa62}
d_s^{\textrm{OSS}} =  0,
\end{equation}
which shows that the conventional OSS scheme achieves a secrecy diversity order of zero. In other words, with a required intercept probability constraint, the outage probability would not approach to zero, as the SNR $\gamma_M$ increases to infinity. Hence, the conventional OSS scheme fails to make the overall intercept probability and overall outage probability both drop to zero with an increasing SNR.

\subsection{Conventional IL-USS}
In this subsection, we analyze the secrecy diversity gain of conventional IL-USS scheme. Similarly to (55), the secrecy diversity of conventional IL-USS scheme can be obtained from (41) as
\begin{equation}\label{equa63}
{d_s^{{\text{IL-USS}}} =  - \mathop {\lim }\limits_{\gamma_M  \to \infty } \frac{{\log \left( {P_{Mm{\text{-out}}}^{{\text{IL-USS}}} \cdot P_{Ss{\text{-out}}}^{{\text{IL-USS}}}} \right)}}{{\log \gamma_M }},}
\end{equation}
where $P_{Mm{\text{-out}}}^{{\text{IL-USS}}}$ and $P_{Ss{\text{-out}}}^{{\text{IL-USS}}}$ are given by (39) and (40), respectively. Denoting $\gamma_S = \beta \gamma_M$ ($0 \le \beta \le \frac{{\sigma _{Mm}^2}}{{\sigma _{Sm}^2}}$) and letting $\gamma_M  \to \infty $, we can simplify $ P_{Me{\text{-int}}}^{{\text{IL-USS}}}$ from (42) as
\begin{equation}\label{equa64}
\mathop {\lim }\limits_{{\gamma _M} \to \infty } P_{Me {\mathop{\textrm{-int}}} }^{{\textrm{IL-USS}}} = \frac{{\sigma _{Me}^2}}{{\sigma _{Me}^2 + \beta \sigma _{Se}^2({2^{R_M^o - R_M^s}} - 1)}},
\end{equation}
which leads to
\begin{equation}\label{equa65}
{2^{R_M^o}} = {2^{R_M^s}}(1 + \frac{{{\Phi^{\textrm{IL-USS}} _{Me}}}}{\beta }),
\end{equation}
where ${\Phi^{\textrm{IL-USS}} _{Me}} = (\frac{1}{{P_{Me {\mathop{\textrm{-int}}} }^{{\textrm{IL-USS}}}}} - 1)\frac{{\sigma _{Me}^2}}{{\sigma _{Se}^2}}$. Using (43) and denoting $\gamma_S = \beta \gamma_M$, we can similarly obtain
\begin{equation}\label{equa66}
{2^{R_S^o}} = {2^{R_S^s}}(1 + \beta{{\Phi^{\textrm{IL-USS}} _{Se}}}),
\end{equation}
for $\gamma_M  \to \infty $, where ${\Phi^{\textrm{IL-USS}} _{Se}} = (\frac{1}{{P_{Se {\mathop{\textrm{-int}}} }^{{\textrm{IL-USS}}}}} - 1)\frac{{\sigma _{Se}^2}}{{\sigma _{Me}^2}}$. Substituting ${2^{R_M^o}}$ and ${2^{R_S^o}}$ from (65) and (66) into (39) and (40) yields
\begin{equation}\label{equa67}
P_{Mm\textrm{-out}}^{{\textrm{IL-USS}}} =\frac{{\beta \sigma _{Sm}^2({2^{R_M^s}} - 1) + {\Phi^{\textrm{IL-USS}} _{Me}}\sigma _{Sm}^2{2^{R_M^s}}}}{{\sigma _{Mm}^2 + \beta \sigma _{Sm}^2({2^{R_M^s}} - 1) + {\Phi^{\textrm{IL-USS}} _{Me}}\sigma _{Sm}^2{2^{R_M^s}}}},
\end{equation}
and
\begin{equation}\label{equa68}
P_{Ss\textrm{-out}}^{{\textrm{IL-USS}}} =\frac{{\sigma _{Ms}^2({2^{R_S^s}} - 1) + \beta \Phi _{Se}^{{\textrm{IL-USS}}}\sigma _{Ms}^2{2^{R_S^s}}}}{{\beta \sigma _{Ss}^2 + \sigma _{Ms}^2({2^{R_S^s}} - 1) + \beta \Phi _{Se}^{{\textrm{IL-USS}}}\sigma _{Ms}^2{2^{R_S^s}}}},
\end{equation}
for $\gamma_M \to \infty$. One can observe from (67) and (68) that as the required intercept probability decreases to zero, the parameters ${\Phi^{\textrm{IL-USS}} _{Me}}$ and ${\Phi^{\textrm{IL-USS}} _{Se}}$ increases to infinity, thus the outage probabilities of macro-cell and small-cell transmissions both increase to one. Combining (63), (67) and (68), we obtain the secrecy diversity gain of conventional IL-USS as
\begin{equation}\label{equa69}
d_s^{\textrm{IL-USS}} =  0,
\end{equation}
from which a secrecy diversity order of zero is achieved by the conventional IL-USS scheme, implying that increasing the transmit power $\gamma_M$ would not force the overall outage probability and intercept probability of conventional IL-USS scheme to be arbitrarily low.

\subsection{Proposed IC-USS}
In this subsection, we present the secrecy diversity analysis for proposed IC-USS scheme. Similarly to (55), the secrecy diversity gain of proposed IC-USS scheme is obtained from (47) as
\begin{equation}\label{equa70}
d_s^{{\text{IC-USS}}} =  - \mathop {\lim }\limits_{\gamma_M  \to \infty } \frac{{\log \left( {P_{Mm{\text{-out}}}^{{\text{IC-USS}}} \cdot P_{Ss{\text{-out}}}^{{\text{IC-USS}}}} \right)}}{{\log \gamma_M }},
\end{equation}
where $P_{Mm{\text{-out}}}^{{\text{IC-USS}}} $ and $P_{Ss{\text{-out}}}^{{\text{IC-USS}}}$ are given by (45) and (46), respectively. Letting $\gamma_M \to \infty$ and using (48), we have
\begin{equation}\label{equa71}
P_{Me{\text{-int}}}^{{\text{IC-USS}}} = \Pr \left(
\begin{split}
&{\frac{{{2^{R_M^s}}(\sigma _{Mm}^2{\beta ^{ - 1}} - \sigma _{Sm}^2)}}{{|{h_{Sm}}{|^2} + |{h_{Mm}}{|^2}|{h_{Se}}{|^2}|{h_{Me}}{|^{ - 2}}}}} \\
&> {2^{R_M^o}} - {2^{R_M^s}}
\end{split}
\right),
\end{equation}
{{from which an arbitrarily small intercept probability can be achieved by the macro-cell transmission relying on our IC-USS scheme through increasing the overall rate ${R_M^o}$. In other words, an overall data rate is always available to satisfy an arbitrary intercept requirement $P_{Me{\text{-int}}}^{{\text{IC-USS}}}$, regardless of $\gamma_M$, which is denoted by ${R_M^o}(P_{Me{\text{-int}}}^{{\text{IC-USS}}})$ for short. Considering ${2^{R_M^o}} \sigma _{Sm}^2 \to 0$ and using (49), the overall rate ${R_M^o}(P_{Me{\text{-int}}}^{{\text{IC-USS}}})$ should satisfy the following equality}}
\begin{equation}\label{equa72}
P_{Me{\textrm{-}}{\mathop{\textrm{int}}} }^{{\textrm{IC-USS}}} - {\Theta _{\lim }}\exp ({\Theta _{\lim }})Ei({\Theta _{\lim }}) = 0,
\end{equation}
where ${\Theta _{\lim }}$ is given by
\begin{equation}\label{equa73}
{\Theta _{\lim }} = \frac{{\sigma _{Mm}^2\sigma _{Me}^2 - \beta \sigma _{Sm}^2\sigma _{Me}^2{2^{R_M^o(P_{Me{\textrm{-}}{\mathop{\textrm{int}}} }^{{\textrm{IC-USS}}}) - R_M^s}}}}{{\beta \sigma _{Mm}^2\sigma _{Se}^2({2^{R_M^o(P_{Me{\textrm{-}}{\mathop{\textrm{int}}} }^{{\textrm{IC-USS}}}) - R_M^s}} - 1)}},
\end{equation}
for $\gamma_M \to \infty$. Meanwhile, letting $\gamma_M  \to \infty $ and ignoring higher-order infinitesimal, we can simplify (45) as
\begin{equation}\label{equa74}
P_{Mm\textrm{-out}}^{\textrm{IC-USS}} = \frac{{{2^{R_M^o}} - 1}}{{\sigma _{Mm}^2 - \sigma _{Sm}^2\beta }} \cdot (\frac{1}{{{\gamma _M}}}).
\end{equation}
Considering an arbitrarily low intercept probability $P_{Me {\mathop{\textrm {-int}}} }^{{\textrm{IC-USS}}} $ and substituting the overall rate ${R_M^o}(P_{Me{\text{-int}}}^{{\text{IC-USS}}})$ into (74) yields
\begin{equation}\label{equa75}
P_{Mm\textrm{-out}}^{\textrm{IC-USS}} = \frac{{{2^{{R_M^o}(P_{Me{\text{-int}}}^{{\text{IC-USS}}})}} - 1}}{{\sigma _{Mm}^2 - \sigma _{Sm}^2\beta }} \cdot (\frac{1}{{{\gamma _M}}}).
\end{equation}
Using (75) and noting the overall rate ${R_M^o}(P_{Me{\text{-int}}}^{{\text{IC-USS}}})$ irrelevant to $\gamma_M$, we can obtain
\begin{equation}\label{equa76}
\mathop {\lim }\limits_{{\gamma _M} \to \infty } \frac{{\log (P_{Mm \textrm{-out}}^{\textrm{IC-USS}})}}{{\log {\gamma _M}}} =  - 1.
\end{equation}

Moreover, letting $\gamma_M \to \infty$, the intercept probability of small-cell transmissions for our IC-USS scheme is obtained from (50) as
\begin{equation}\label{equa77}
P_{Se{\textrm{-int}}}^{{\textrm{IC-USS}}} = \Pr \left(
\begin{split}
&{\frac{{|{h_{Se}}{|^2}{{\left| {{h_{Mm}}} \right|}^2}}}{{|{h_{Me}}{|^2}(\sigma _{Mm}^2 + \beta |{h_{Sm}}{|^2} - \beta \sigma _{Sm}^2)}}}\\
&> \frac{{({2^{R_S^o - R_S^s}} - 1)}}{\beta }
\end{split}
\right),
\end{equation}
{{which shows an overall data rate as denoted by ${R_S^o}(P_{Se{\text{-int}}}^{{\text{IC-USS}}})$ is available to guarantee an arbitrary small intercept probability $P_{Me{\text{-int}}}^{{\text{IC-USS}}}$, regardless of $\gamma_M$. Considering ${2^{R_S^o}} \sigma _{Sm}^2 \to 0$ and using (51), the overall rate ${R_S^o}(P_{Se{\text{-int}}}^{{\text{IC-USS}}})$ can be obtained by solving the following equality}}
\begin{equation}\label{equa78}
1 - \Psi_{Se} \exp (\Psi_{Se} )Ei(\Psi_{Se} ) - P_{Se{\textrm{-int}}}^{{\textrm{IC-USS}}} = 0,
\end{equation}
where $\Psi_{Se}$ is given by
\begin{equation}\nonumber
\Psi_{Se}  = \frac{{\sigma _{Me}^2{2^{R_S^o(P_{Se{\textrm{-int}}}^{{\textrm{IC-USS}}})}} - \sigma _{Me}^2{2^{R_S^s}}}}{{\beta \sigma _{Se}^2{2^{R_S^s}}}}.
\end{equation}
Similarly, letting $\gamma_M \to \infty$, we can obtain the outage probability of small-cell transmissions for our IC-USS scheme from (25) as
\begin{equation}\label{equa79}
P_{Ss{\text{-out}}}^{{\text{IC-USS}}}=\Pr \left(
\begin{split}
&{\frac{{|{h_{Ss}}{|^2}{{\left| {{h_{Mm}}} \right|}^2}}}{{|{h_{Ms}}{|^2}(\sigma _{Mm}^2 + \beta |{h_{Sm}}{|^2} - \beta \sigma _{Sm}^2)}}}\\
&< \frac{({2^{R_S^o}} - 1)}{\beta}
\end{split}
\right),
\end{equation}
{{from which an outage probability floor is converged with a target overall data rate ${R_S^o}(P_{Se{\text{-int}}}^{{\text{IC-USS}}})$ of guaranteeing a required intercept probability. This means that with an arbitrary small intercept probability requirement, it is impossible to achieve an arbitrary low outage probability for the small cell relying on our IC-USS scheme. Considering ${2^{R_S^o}} \sigma _{Sm}^2 \to 0$ and using (46), we have}}
\begin{equation}\label{equa80}
P_{Ss{\textrm{-out}}}^{{\textrm{IC-USS}}} = {\Psi _{Ss}}\exp ({\Psi _{Ss}})Ei({\Psi _{Ss}}),
\end{equation}
for $\gamma_M \to \infty$, where ${\Psi _{Ss}} = \sigma _{Ms}^2({2^{R_S^o}} - 1)/(\beta \sigma _{Ss}^2)$. It can be observed from (80) that with a target overall data rate ${R_S^o}(P_{Se{\text{-int}}}^{{\text{IC-USS}}})$, the outage probability would not be arbitrarily small, as the SNR $\gamma_M$ approaches to the infinity, which, in turn, leads to
\begin{equation}\label{equa81}
\mathop {\lim }\limits_{{\gamma _M} \to \infty } \frac{{\log (P_{Ss \textrm{-out}}^{\textrm{IC-USS}})}}{{\log {\gamma _M}}} =  0.
\end{equation}
Finally, by combining (76) and (81) with (70), the secrecy diversity gain of proposed IC-USS scheme is given by
\begin{equation}\label{equa82}
\begin{split}
d_s^{\textrm{IC-USS}} = 1,
\end{split}
\end{equation}
which indicates that the proposed IC-USS scheme can achieve the secrecy diversity order of one higher than both the conventional OSS and IL-USS schemes. This also implies that with an arbitrarily low overall intercept probability, the overall outage probability of proposed IC-USS scheme would asymptotically decrease to zero, as the SNR $\gamma_M$ increases to infinity. As a consequence, with an increasing SNR of $\gamma_M$, the proposed IC-USS scheme can make the overall outage probability and overall intercept probability both drop to zero, showing its significant advantage over the conventional OSS and IL-USS methods from an SRT perspective.

\section{Numerical Results And Discussions}
In this section, we present numerical performance comparisons among the conventional OSS and IL-USS methods as well as the proposed IC-USS scheme in terms of their overall outage probability and intercept probability. In our numerical evaluation, we consider $\sigma _{Mm}^2 = \sigma _{Ss}^2 = \sigma _{Me}^2 = \sigma _{Se}^2 = 1$, in which the eavesdropper is assumed to experience the same fading gain as the MU and SU. Since a small cell may be deployed in a shadowed area (e.g., underground parking garage, tunnel, etc.) of the macro cell, a fading gain of $\sigma _{Ms}^2 = \sigma _{Sm}^2 = 0.1$ is considered for interference channels between the small cell and macro cell. Also, an SNR of $\gamma_M = 25{\textrm{ dB}}$, a secrecy data rate of $R^s_M = R^s_S = 0.5{\textrm{ bit/s/Hz}}$ and $\alpha=\beta=0.5$ are used, unless otherwise stated. Additionally, simulated intercept probability and outage probability results of OSS, IL-USS and IC-USS schemes are also provided through Monte-Carlo simulation to verify our theoretical SRT analysis, where theoretical SRT results are obtained by plotting (37), (38), (41), (44), (47) and (52).

\begin{figure}
\centering
\includegraphics[scale=0.55]{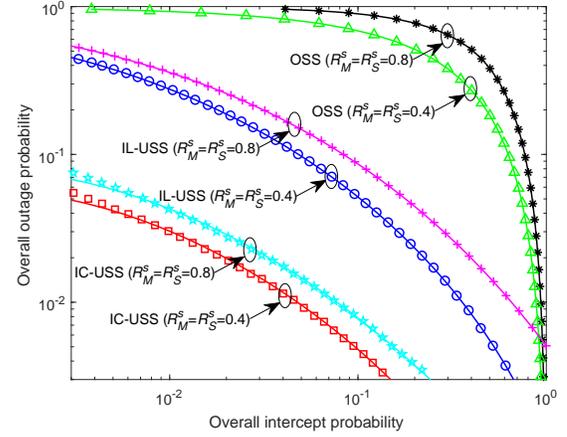}
\caption{Overall outage probability versus overall intercept probability of OSS, IL-USS and IC-USS schemes for different secrecy rates of $R^s_M = R^s_S = 0.4$ and $0.8{\textrm{ bit/s/Hz}}$ by adjusting the overall data rates of $R^o_M = R^o_S$ in the range of $[R^s_M,5]$, where solid lines and discrete markers represent the theoretical and simulation results, respectively.}
\label{fig 2}
\end{figure}
Fig. 2 shows the overall outage probability versus overall intercept probability of the conventional OSS and IL-USS methods as well as the proposed IC-USS scheme for different secrecy rates of $R^s_M = R^s_S = 0.4$ and $0.8{\textrm{ bit/s/Hz}}$, where solid lines and discrete markers represent the theoretical and simulation results, respectively. As observed in Fig. 2, with an increasing overall intercept probability, the overall outage probabilities of OSS, IL-USS and IC-USS schemes decrease and vice versa, showing a tradeoff between the security and reliability, called security-reliability tradeoff (SRT). Fig. 2 also shows that as the secrecy rate increases from $R^s_M = R^s_S = 0.4 $ to $ 0.8{\textrm{ bit/s/Hz}}$, the SRT performance of OSS, IL-USS and IC-USS degrades accordingly, and the proposed IC-USS scheme performs better than the conventional OSS and IL-USS methods. Moreover, the theoretical and simulated SRT results of Fig. 2 generally match well, verifying the correctness of our derived closed-form expressions of overall outage probability and intercept probability. It is noted that when the overall intercept probability is small, the gap between the theoretical and simulated overall outage probabilities of proposed IC-USS becomes obvious. This is because that with a continuously decreasing intercept probability, the overall data rates of $R_M^o$ and $R_S^o$ increase to be sufficiently high such that the asymptotic assumptions of ${2^{R_M^o}} \sigma _{Sm}^2 \to 0$ and ${2^{R_S^o}} \sigma _{Sm}^2 \to 0$ become no longer valid, resulting in an accuracy loss of our derived outage probability and intercept probability expressions of (46), (49) and (51) for IC-USS scheme.

\begin{figure}
\centering
\includegraphics[scale=0.55]{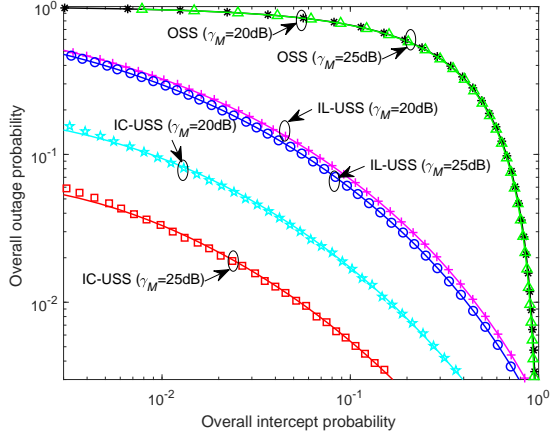}
\caption{Overall outage probability versus overall intercept probability of OSS, IL-USS and IC-USS schemes for different SNRs of $\gamma_M = 20$ and $25{\textrm{ dB}}$ by adjusting the overall data rates of $R^o_M = R^o_S$ in the range of $[R^s_M,5]$, where solid lines and discrete markers represent the theoretical and simulation results, respectively.}
\label{fig 3}
\end{figure}
Fig. 3 depicts the overall outage probability versus overall intercept probability of OSS, IL-USS and IC-USS schemes for different SNRs of $\gamma_M = 20 $ and $25{\textrm{ dB}}$. One can observe from Fig. 3 that as the SNR $\gamma_M$ increases from $20 $ to $ 25{\textrm{ dB}}$, the SRT performance of IL-USS and IC-USS improves accordingly, whereas no improvement is achieved by the OSS scheme. This is because that although increasing the transmit power improves the transmission reliability in terms of decreasing the outage probability of a legitimate receiver, it also results in an improved signal reception at the eavesdropper along with a degraded security performance. Overall speaking, no SRT benefit is attained by the OSS scheme, considering the eavesdropper having the same fading gain as the MU and SU (i.e., $\sigma _{Mm}^2 = \sigma _{Ss}^2 = \sigma _{Me}^2 = \sigma _{Se}^2 = 1$). By contrast, in the IL-USS and IC-USS approaches, the macro cell and small cell interfere with each other and increasing the transmit power would cause higher interference to the eavesdropper as well as the MU and SU, leading to a secrecy improvement at the cost of degrading the reliability. Moreover, with an increased transmit power of $P_M$, the eavesdropper encounters more interference than the legitimate MU and SU, since the interference channel gains of $\sigma _{Ms}^2 = \sigma _{Sm}^2 = 0.1$ are much smaller than the eavesdropping channel gains of $\sigma _{Me}^2 = \sigma _{Se}^2 = 1$. Therefore, as the SNR $\gamma_M$ increases from $ 20 $ to $ 25{\textrm{ dB}}$, the secrecy improvement dominates over the reliability degradation for both IL-USS and IC-USS schemes, leading to their SRT improvements. Additionally, as seen from Fig. 3, with an increased SNR of $\gamma_M$, the SRT enhancement of proposed IC-USS is much more significant than that of IL-USS, which is due to the fact that our IC-USS scheme is sophisticatedly designed to alleviate the interference problem by canceling out the SBS-MU interference.

\begin{figure}
\centering
\includegraphics[scale=0.55]{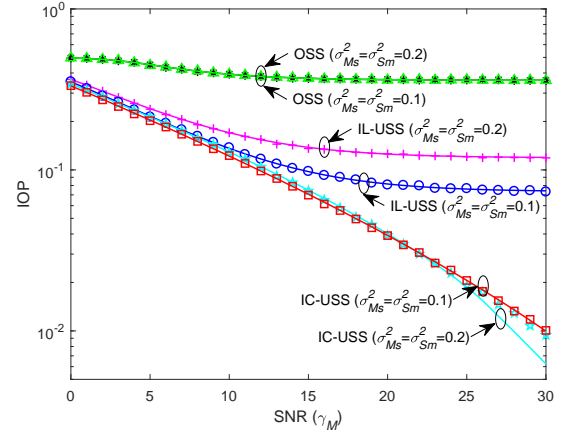}
\caption{Intercept and outage probability (IOP) versus SNR $\gamma_M$ of OSS, IL-USS and IC-USS schemes, where solid lines and discrete markers represent the theoretical and simulation results, respectively.}
\label{fig 4}
\end{figure}
In Fig. 4, we show the sum of overall intercept probability and outage probability (as denoted by IOP for short) versus the SNR $\gamma_M$ of OSS, IL-USS and IC-USS schemes for different interference channel gains of $\sigma _{Ms}^2 = \sigma _{Sm}^2 = 0.1$ and $0.2$. It needs to be pointed out that given an SNR $\gamma_M$, the numerical IOP results of OSS, IL-USS and IC-USS schemes are minimized by adjusting the overall data rates of $R^o_M $ and $ R^o_S$. As shown in Fig. 4, with an increasing SNR $\gamma_M$, the IOP of conventional OSS method almost keeps unchanged, further verifying that no SRT improvement is obtained by the OSS scheme in the case of $\sigma _{Mm}^2 = \sigma _{Ss}^2 = \sigma _{Me}^2 = \sigma _{Se}^2 = 1$. Fig. 4 also shows that as the SNR $\gamma_M$ increases, the IOP of IL-USS scheme initially decreases and then converges to a floor, while the proposed IC-USS scheme continuously decreases the IOP significantly. It can be observed from Fig. 4 that with an increasing SNR $\gamma_M$, the performance advantage of proposed IC-USS scheme over conventional OSS and IL-USS methods becomes more significant in terms of their IOPs. In addition, as the interference channel gains of $\sigma _{Ms}^2$ and $ \sigma _{Sm}^2 $ increase from $0.1$ to $0.2$, the IOP performance of IL-USS worsens substantially, whereas our IC-USS scheme is very resistant to the interference channels without an IOP degradation. Fig. 4 further demonstrates an obvious gap between the theoretical and simulated IOP results in the high SNR region for our IC-USS scheme in the case of $\sigma _{Ms}^2 = \sigma _{Sm}^2 = 0.2$, which is because that the asymptotic assumptions of ${2^{R_M^o}} \sigma _{Sm}^2 \to 0$ and ${2^{R_S^o}} \sigma _{Sm}^2 \to 0$ used in our theoretical SRT analysis for IC-USS become invalid, as the interference channel gain of $ \sigma _{Sm}^2 $ increases.

\begin{figure}
\centering
\includegraphics[scale=0.55]{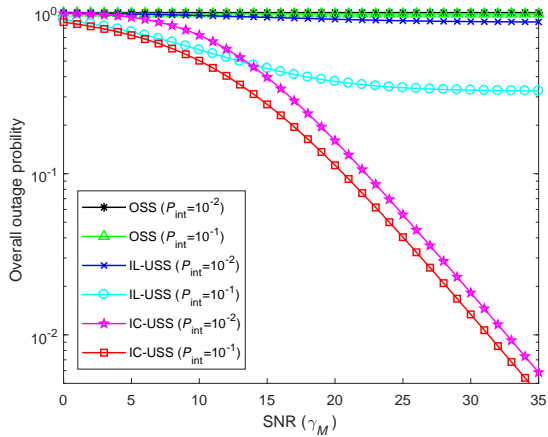}
\caption{Overall outage probability versus SNR $\gamma_M$ of OSS, IL-USS and IC-USS schemes for different individual intercept probability constraints of $P_{\textrm{int}} =0.1$ and $ 0.01$ for both the macro-cell and small-cell transmissions.}
\label{fig 5}
\end{figure}
Fig. 5 illustrates the overall outage probability versus SNR $\gamma_M$ of OSS, IL-USS and IC-USS schemes for different individual intercept probability constraints of $P_{\textrm{int}} = 10^{-1}$ and $P_{\textrm{int}} = 10^{-2}$, where the intercept probabilities of macro-cell and small-cell transmissions each shall be less than a required level. As shown in Fig. 5, as the intercept probability requirement relaxes from $10^{-2}$ to $10^{-1}$, the overall outage probabilities of OSS, IL-USS and IC-USS schemes all decrease and the proposed IC-USS scheme achieves the best outage performance with a given SNR. Additionally, as the SNR $\gamma_M$ increases, the conventional OSS and IL-USS methods converge to their respective outage probability floors, whereas the overall outage performance of proposed IC-USS scheme continuously improves without the floor effect.

\begin{figure}
\centering
\includegraphics[scale=0.55]{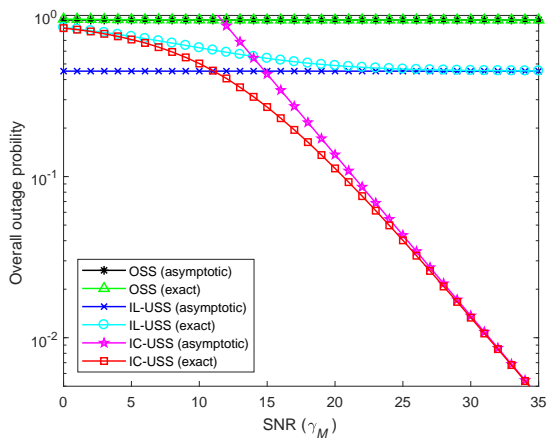}
\caption{Exact and asymptotic overall outage probabilities versus SNR $\gamma_M$ of OSS, IL-USS and IC-USS schemes with a required individual intercept probability constraint of $P_{\textrm{int}} = 0.05$.}
\label{fig 6}
\end{figure}
In Fig. 6, we show the exact and asymptotic overall outage probabilities versus SNR of OSS, IL-USS and IC-USS schemes with a required intercept probability constraint, where the asymptotic overall outage probabilities are obtained by using (60), (61), (67), (68), (72), (75), (78) and (80). It is shown from Fig. 6 that the asymptotic outage probability curves of OSS, IL-USS and IC-USS schemes converge to their respective exact outage results, as the SNR $\gamma_M$ increases. This is because that the derivation of our asymptotic SRT expressions takes into account the assumption of $\gamma_M \to \infty$. One can also observe from Fig. 6 that with an increasing SNR $\gamma_M$, the proposed IC-USS scheme can make its overall outage probability decrease significantly with a required intercept probability constraint. This means that the SRT performance of proposed IC-USS scheme can be continuously improved by simply increasing the transmit power.

\section{Concluding Remarks}
In this paper, we investigated physical-layer security for a heterogeneous spectrum-sharing cellular network consisting of a macro cell and a small cell in the presence of a common eavesdropper. We proposed an IC-USS scheme to improve transmission security of the heterogeneous cellular network against eavesdropping, where both the macro cell and small cell are allowed to simultaneously transmit over their shared spectrum along with mutual interference induced between each other. A special signal was designed in the proposed IC-USS scheme to alleviate an adverse effect of the mutual interference on an intended user while severely degrading the eavesdropper. Conventional OSS and IL-USS methods were considered for the purpose of performance comparisons. We derived closed-form expressions of overall outage probability and intercept probability for the OSS, IL-USS and IC-USS schemes. We also conducted the secrecy diversity analysis for OSS, IL-USS and IC-USS schemes and showed that the proposed IC-USS scheme achieves a higher secrecy diversity gain than the OSS and IL-USS methods. Additionally, numerical results demonstrated that the IC-USS scheme performs better than the conventional OSS and IL-USS methods in terms of their SRT performance. More importantly, with an arbitrarily low intercept probability, the overall outage probabilities of conventional OSS and IL-USS methods converge to their respective outage floors, whereas the proposed IC-USS scheme can make its overall outage probability asymptotically decrease to zero, as the SNR increases to infinity.

\begin{IEEEbiography}[{\includegraphics[width=1in,height=1.25in]{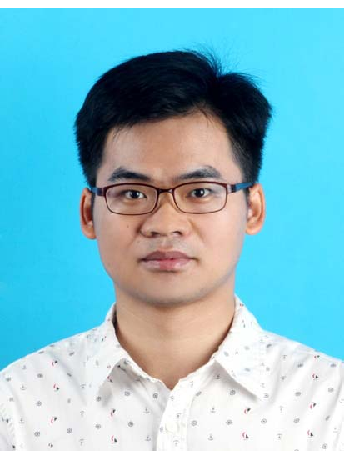}}]{Yulong Zou} (SM'13) is a Full Professor and Doctoral Supervisor at the Nanjing University of Posts and Telecommunications (NUPT), Nanjing, China. He received the B.Eng. degree in information engineering from NUPT, Nanjing, China, in July 2006, the first Ph.D. degree in electrical engineering from the Stevens Institute of Technology, New Jersey, USA, in May 2012, and the second Ph.D. degree in signal and information processing from NUPT, Nanjing, China, in July 2012.

Dr. Zou was awarded the 9th IEEE Communications Society Asia-Pacific Best Young Researcher in 2014 and a co-receipt of the Best Paper Award at the 80th IEEE Vehicular Technology Conference in 2014. He has served as an editor for the IEEE Communications Surveys \& Tutorials, IEEE Communications Letters, EURASIP Journal on Advances in Signal Processing, IET Communications, and China Communications. In addition, he has acted as TPC members for various IEEE sponsored conferences, e.g., IEEE ICC/GLOBECOM/WCNC/VTC/ICCC, etc.

\end{IEEEbiography}


\begin{thebibliography}{1}

\bibitem{IEEEhowto:1}
S. Chen, F. Qin, B. Hu, X. Li, and Z. Chen, ``User-centric ultra-dense networks (UUDN) for 5G: Challenges, methodologies and directions," \emph{IEEE Wirel. Commun.}, vol. 23, no. 2, pp. 78-85, Apr. 2016.

\bibitem{IEEEhowto:2}
C. Wang, F. Haider, X. Gao, \emph{et al.}, ``Cellular architecture and key technologies for 5G wireless communication networks," \emph{IEEE Commun. Mag.}, vol. 52, no. 2, pp. 122-130, Feb. 2014.

\bibitem{IEEEhowto:3}
S. Chen, B. Ren, Q. Gao, \emph{et al.}, ``Pattern division multiple access (PDMA) - A novel non-orthogonal multiple access for fifth-generation radio networks," \emph{IEEE Trans. Veh. Tech.}, vol. 66, no. 4, pp. 3185-3196, Apr. 2017.

\bibitem{IEEEhowto:4}
R. Madan, J. Borran, A. Sampath, \emph{et al.}, ``Cell association and interference coordination in heterogeneous LTE-A cellular networks," \emph{IEEE J. Sel. Areas Commun.}, vol. 28, no. 9, pp. 1479-1489, Spet. 2010.

\bibitem{IEEEhowto:5}
S. Yong, T. Quek, M. Kountouris, and G. Caire, ``Cognitive hybrid division duplex for two-tier femtocell networks," \emph{IEEE Trans. Wirel. Commun.}, vol. 12, no. 10, pp. 4852-4865, Oct. 2013.

\bibitem{IEEEhowto:6}
C. Yang, J. Li, A. Anpalagan, and M. Guizani, ``Joint power coordination for spectral-and-energy efficiency in heterogeneous small cell networks: A bargaining game-theoretic perspective," \emph{IEEE Trans. Wirel. Commun.}, vol. 15, no. 2, pp. 1364-1376, Feb. 2016.

\bibitem{IEEEhowto:7}
X. Li, T. Jiang, S. Cui, J. An, and Q. Zhang, ``Cooperative communications based on rateless network coding in distributed MIMO systems," \emph{IEEE Wirel. Commun.}, vol. 17, no. 3, pp. 60-67, Jun. 2010.

\bibitem{IEEEhowto:8}
F. Martin-Vega, G. Gomez, M. Aguayo-Torres, \emph{et al.}, ``Analytical modeling of interference aware power control for the uplink of heterogeneous cellular networks," \emph{IEEE Trans. Wirel. Commun.}, vol. 15, no. 10, pp. 6742-6757, Oct. 2016.

\bibitem{IEEEhowto:9}
H. Hu, H. Wang, Q. Zhu, \emph{et al.}, ``Uplink performance analysis in multi-tier heterogeneous cellular networks with power control and biased user association," \emph{China Commun.}, vol. 13, no. 12, pp. 25-36, Dec. 2016.

\bibitem{IEEEhowto:10}
Z. Yu, K. Wang, H. Ji, \emph{et al.}, ``Dynamic resource allocation in TDD-based heterogeneous cloud radio access networks," \emph{China Commun.}, vol. 13, no. 6, pp. 1-11, Jun. 2016.

\bibitem{IEEEhowto:11}
K. Wang, H. Li, F. Yu, \emph{et al.}, ``Interference alignment in virtualized heterogeneous cellular networks with imperfect channel state information," \emph{IEEE Trans. Veh. Tech.}, vol. 66, no. 2, pp.1519-1532, Feb. 2017.

\bibitem{IEEEhowto:12}
A. He, L. Wang, Y. Chen, K. K. Wong, and M. Elkashlan, ``Uplink interference management in massive MIMO enabled heterogeneous cellular networks," \emph{IEEE Wirel. Commun. Lett.}, vol. 5, no.5, pp. 560-563, May 2017.

\bibitem{IEEEhowto:13}
Y. Zou, J. Zhu, X. Wang, and L. Hanzo, ``A survey on wireless security: Technical challenges, recent advances and future trends," \emph{Proc. of the IEEE}, vol. 104, no. 9, pp. 1727-1765, Sept. 2016.

\bibitem{IEEEhowto:14}
M. Bloch and J. Barros, ``\emph{Physical-layer security: From information theory to security engineering}," UK: Cambridge University Press, 2011.

\bibitem{IEEEhowto:15}
Y. Zou, J. Zhu, X. Wang, and V. C. M. Leung, ``Improving physical-layer security in wireless communications using diversity techniques," \emph{IEEE Net.}, vol. 29, no. 1, pp. 42-48, Feb. 2015.

\bibitem{IEEEhowto:16}
A. D. Wyner, ``The wire-tap channel," \emph{Bell Syst. Tech. J.}, vol. 54, no. 8, pp. 1355-1387, Aug. 1975.

\bibitem{IEEEhowto:17}
S. K. Leung-Yan-Cheong and M. E. Hellman, ``The Gaussian wiretap channel," \emph{IEEE Trans. Inf. Theory}, vol. 24, pp. 451-456, Jul. 1978.

\bibitem{IEEEhowto:18}
A. Khisti and D. Zhang, ``Artificial-noise alignment for secure multicast using multiple antennas," \emph{IEEE Commun. Lett.}, vol. 17, no. 8, pp. 1568-1571, Aug. 2013.

\bibitem{IEEEhowto:19}
Y. Wu, R. Schober, D. Ng, C. Xiao, and G. Caire, ``Secure massive MIMO transmission with an active eavesdropper," \emph{IEEE Trans. Inf. Theory}, vol. 62, no. 7, pp. 3880-3900, Jul. 2016.

\bibitem{IEEEhowto:20}
Y. Wu, C. Xiao, Z. Ding, X. Gao, and S. Jin, ``Linear precoding for finite-alphabet signaling over MIMOME wiretap channels," \emph{IEEE Trans. Veh. Tech.}, vol. 61, no. 6, pp. 2599-2612, Jun. 2012.

\bibitem{IEEEhowto:21}
J. Zhu, Y. Zou, G. Wang, Y. D. Yao, and G. K. Karagiannidis, ``On secrecy performance of antenna-selection-aided MIMO systems against eavesdropping," \emph{IEEE Trans. Veh. Technol.}, vol. 65, no. 1, pp. 214-225, Jan. 2016.

\bibitem{IEEEhowto:22}
Y. Zou, X. Wang, and W. Shen, ``Physical-layer security with multiuser scheduling in cognitive radio networks", IEEE Transactions on Communications, vol. 61, no. 12, pp. 5103 - 5113, Dec. 2013.

\bibitem{IEEEhowto:23}
Y. Zou, X. Wang, and W. Shen, ``Optimal relay selection for physical-layer security in cooperative wireless networks," \emph{IEEE J. Sel. Areas Commun.}, vol. 31, no. 10, pp. 2099-2111, Oct. 2013.

\bibitem{IEEEhowto:24}
G. Zheng, L. Choo, and K. K Wong, ``Optimal cooperative jamming to enhance physical layer security using relays," \emph{IEEE Trans. Signal Process.}, vol. 59, no. 3, pp. 1317-1322, Mar. 2011.

\bibitem{IEEEhowto:25}
H. Guo, Z. Yang, L. Zhang, J. Zhu, and Y. Zou, ``Optimal power allocation for joint relay and jammer selection assisted wireless physical-layer security," \emph{IEEE Trans. Commun.}, vol. 65. No. 5, pp. 2180-2193, May 2017.

\bibitem{IEEEhowto:26}
Y. Zou, ``Physical-layer security for spectrum sharing systems," \emph{IEEE Trans. Wirel. Commun.}, vol. 16, no. 2, pp. 1319-1329, Feb. 2017.

\bibitem{IEEEhowto:27}
Y. Zou, J. Zhu, L. Yang, Y.-.C. Liang, and Y.-D. Yao, ``Securing physical-layer communications for cognitive radio networks," \emph{IEEE Commun. Mag.}, vol. 53, no. 9, pp. 48-54, Sept. 2015.

\bibitem{IEEEhowto:28}
W. Wang, K. Teh, and K. Li, ``Enhanced physical layer security in D2D spectrum sharing networks," \emph{IEEE Wirel. Commun. Lett.}, vol. 6, no. 1, pp. 106-109, Jan. 2017.

\bibitem{IEEEhowto:29}
Y. Pei, Y.-C. Liang, K. C. Teh, and K. Li, ``Secure communication in multiantenna cognitive radio networks with imperfect channel state information," \emph{IEEE Trans. Signal Process.}, vol. 59, no. 4, pp. 1683-1693, Apr. 2011.

\bibitem{IEEEhowto:30}
Y. Zou, X. Li, and Y. C. Liang, \textquotedblleft Secrecy outage and diversity analysis of cognitive radio systems," \emph {IEEE J. Sel. Areas Commun.},vol. 32, no. 11, pp. 2222-2236, Nov. 2014.

\bibitem{IEEEhowto:31}
Y. Zou, B. Champagne, W.-P. Zhu, and L. Hanzo, ``Relay-selection improves the security-reliability trade-off in cognitive radio systems," \emph{IEEE Trans. Commun.}, vol. 63, no. 1, pp. 215-228, Jan. 2015.

\bibitem{IEEEhowto:32}
C. Ma, J. Liu, X. Tian, \emph{et al.}, ``Interference exploitation in D2D-enabled cellular networks: A secrecy perspective," \emph{IEEE Trans. Commun.}, vol. 63, no. 1, pp. 229-242, Jan. 2015.

\bibitem{IEEEhowto:33}
R. Zhang, X. Cheng, and L. Yang, \textquotedblleft Cooperation via spectrum sharing for physical layer security in device-to-device communications underlaying cellular networks," \emph{IEEE Trans. Wirel. Commun.}, vol. 15, no. 8, pp. 5651-5663, Aug. 2016.

\bibitem{IEEEhowto:34}
{{T.-X. Zheng, H.-M. Wang, Q. Yang, and M. H. Lee, ``Safeguarding decentralized wireless networks using full-duplex jamming receivers," \emph{IEEE Trans. Wirel. Commun.}, vol. 16, no. 1, pp. 278-292, Jan. 2017.}}

\bibitem{IEEEhowto:35}
{{T.-X. Zheng, H.-M. Wang, J. Yuan, Z. Han, and M. H. Lee, ``Physical layer security in wireless ad hoc networks under a hybrid full-/half-duplex receiver deployment strategy," \emph{IEEE Trans. Wirel. Commun.}, vol. 16, no. 6, pp. 3827-3839, Mar. 2017.}}

\bibitem{IEEEhowto:36}
{{Y. Sun, N. Ng, J. Zhu, \emph{et al.}, ``Robust and secure resource allocation for full-duplex MISO multicarrier NOMA systems," online available: https://arxiv.org/abs/1710.01391, 2017.}}

\bibitem{IEEEhowto:37}
G. Wang, Q. Liu, R. He, and F. Gao, ``Acquisition of channel state information in heterogeneous cloud radio access networks: Challenges and research directions," \emph{IEEE Wirel. Commun.}, vol. 22, no. 3, pp. 100-107, Jun. 2015.

\bibitem{IEEEhowto:38}
Y. Zou, J. Zhu, X. Li, and L. Hanzo, ``Relay selection for wireless communications against eavesdropping: A security-reliability tradeoff perspective", \emph{IEEE Net.}, vol. 30, no. 5, pp. 74-79, Sept. 2016.

\bibitem{IEEEhowto:39}
X. Tang, R. Liu, P. Spasojevic, and H. V. Poor, ``On the throughput of secure hybrid-ARQ protocols for Gaussian block-fading channels," \emph{IEEE Trans. Inf. Theory}, vol. 55, no. 4, pp. 1575-1591, Apr. 2009.

\end{thebibliography}
\end{document}